\documentclass{appolb}
\usepackage{graphicx}

\begin{document}

\title{Scaling behavior of transverse momenta distributions in
hadronic and nuclear collisions%
\thanks{Presented at the 42 International Symposium on Multiparticle
Dynamics (ISMD 2012), Kielce, Poland.} }
\author{M. Rybczy\'{n}ski\thanks{e-mail: maciej.rybczynski@ujk.edu.pl},
Z. W\l{}odarczyk\thanks{e-mail: zbigniew.wlodarczyk@ujk.edu.pl}
\address{Institute of Physics, Jan Kochanowski University, \'Swietokrzyska 15,
25-406 Kielce, Poland}
\and {Grzegorz Wilk\thanks{e-mail: wilk@fuw.edu.pl}}
\address{National Centre for Nuclear Research, Ho\.{z}a 69, 00-681 Warsaw, Poland}
}
\maketitle
\begin{abstract}
It has been recently noticed that transverse momenta ($p_T$)
distributions observed in high energy production processes exhibit
remarkably universal scaling behavior. This is seen when they are
in some suitable variable, replacing the usual $p_T$. On the other
hand, it is also known that transverse momenta distributions in
general follow a power-like Tsallis distribution, rather than an
exponential Boltzmann-Gibbs, with a (generally energy dependent)
nonextensivity parameter $q$. We now show that it is possible to
choose a suitable variable such that $p_T$ distributions of
particles produced in proton-proton interactions in a wide energy
range can be fitted by the same Tsallis distribution (with the
same, energy independent, value of the $q$-parameter). Similar
scaling behavior in nucleus-nucleus collisions is also observed.
The possible dynamical origin of the q parameter used in these
fits will be discussed.
\end{abstract}

\vspace{1cm}

It is well known \cite{WW,cond,Others} that $p_T$ distributions in
general follow a two-parameter power-like Tsallis distribution
\cite{Tsallis} characterized by some energy dependent
nonextensivity parameter $q$, which for $q \rightarrow 1$ becomes
the usual one-parameter exponential Boltzmann-Gibbs (BG)
distribution\footnote{In phenomenological fits the Tsallis
distribution is sometimes used interchangeably with the old two
parameter, purely phenomenological, power-like parametrization,
the so called "Hagedorn formula" \cite{Oldpar}. Here we shall not
discuss this possibility.}:
\begin{equation}
h_q\left( p_T\right) = C_q\cdot\left[ 1 -
(1-q)\frac{p_T}{T}\right]^{\frac{1}{1-q}} \quad \stackrel{q
\rightarrow 1}{\Longrightarrow}\quad h\left( p_T\right) =
C_1\cdot\exp \left(-\frac{p_T}{T}\right),\label{eq:Tsallis}
\end{equation}
(where $C_q$ is a normalization constant)\footnote{Actually, as
shown in \cite{CYWW}, $h_q\left( p_T\right)$ can fit {\it all}
available data on $p_T$ distributions from RHIC to LHC, i.e., up
to $p_T ~200$ GeV/c. The fact that data behave in such a way that
they can be fitted by a simple two-parameter Tsallis formula for
such a broad $p_T$ range is really a phenomenon awaiting a proper
understanding.}. The transverse momenta distributions are then
supposed to bring information on the thermodynamical properties of
the production process, on the temperature $T$ parameter in the
case of exponential  BG distributions, or, additionally, also on
some intrinsic fluctuations presented in such systems and
described by the nonextensivity parameter $q$ in Tsallis
distributions.

It was recently shown that transverse momenta ($p_T$)
distributions observed in high energy production processes exhibit
a universal scaling behavior when presented in suitable variables,
for example: \vspace{-0.5cm}
\begin{figure}[htb]
\includegraphics[width=6.5cm]{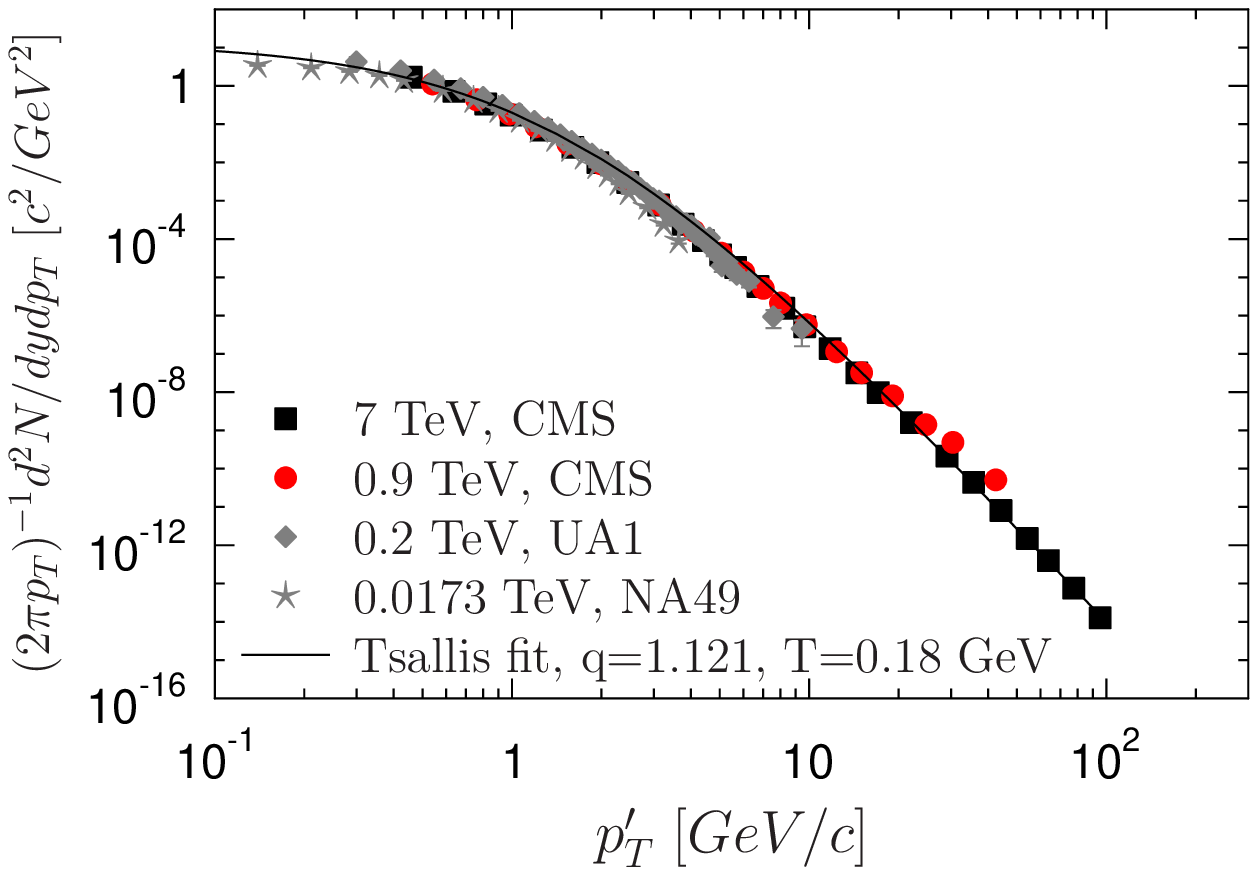}
\includegraphics[width=6.5cm]{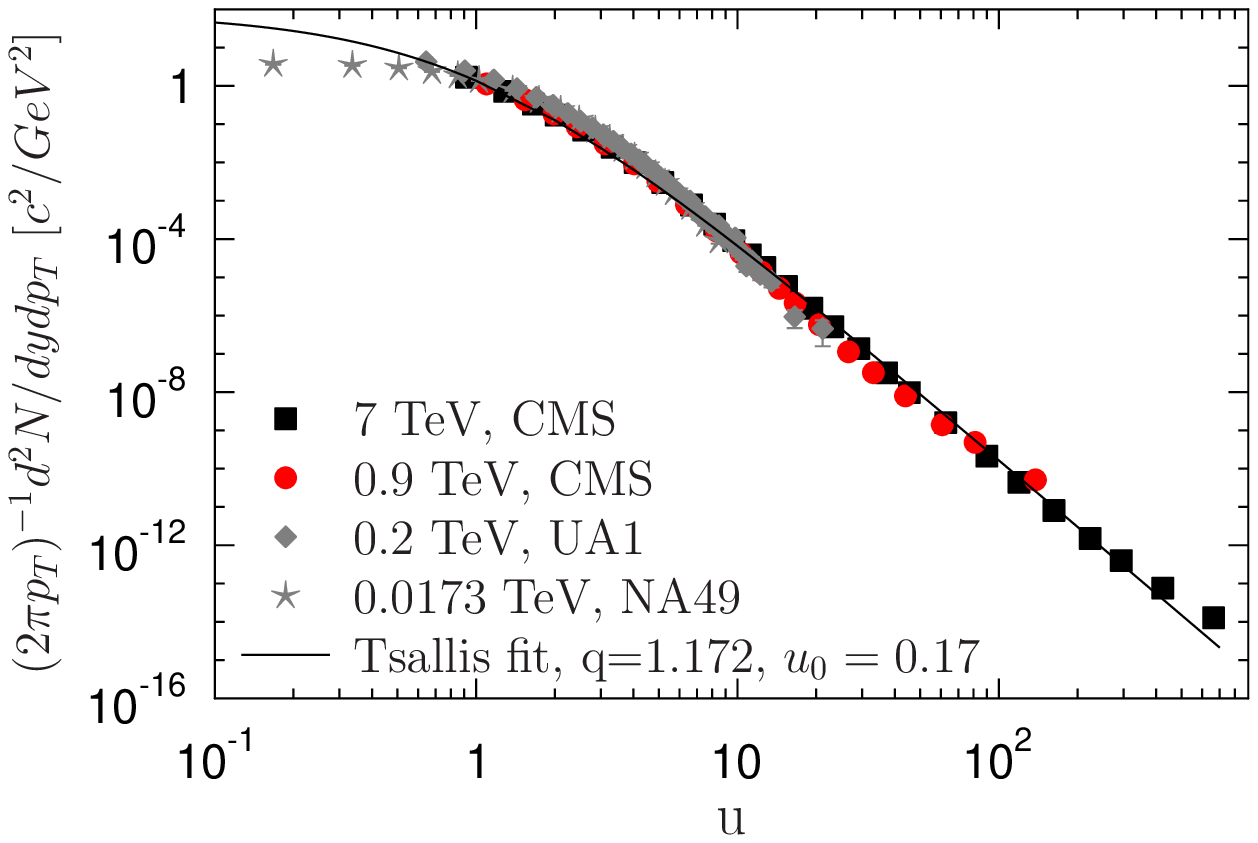}
\caption{(Color online) $p+p$, ($p+\bar{p}$) data for transverse
momentum distributions for different energies \cite{CMS,UA1,NA49}
plotted by using as scaling variables $p'_T$ (left panel) and $u$
(right panel).} \label{Fig1}
\end{figure}
\begin{itemize}
\item The variable $p'_T$ \cite{MP}. It is defined by demanding
that $p_T$ at energy $W$ should be connected with $p'_T$ at energy
$W'$ via $ p'_T = p_T\left( W'/W\right)^{\frac{\lambda}{\lambda +
2}}$; it can be reproduced by a two-parameter Tsallis fit,
$h_q\left( p_t'\right)$ , cf. Fig. \ref{Fig1} (left panel);

\item The variable $u=u\left(p_T\right) = \frac{p_T}{\langle
p_T\rangle - b\cdot p_T}$. It is discussed in \cite{RWW}, cf. Fig.
\ref{Fig1} (right panel) showing $h_q(u)$; here $b$ is an energy
dependent parameter, $b=b(s)$.
\end{itemize}
In both cases $q$ is energy independent, i.e., one obtains a kind
of $q$-scaling phenomenon.

\begin{figure}[htb]
\includegraphics[width=6.5cm]{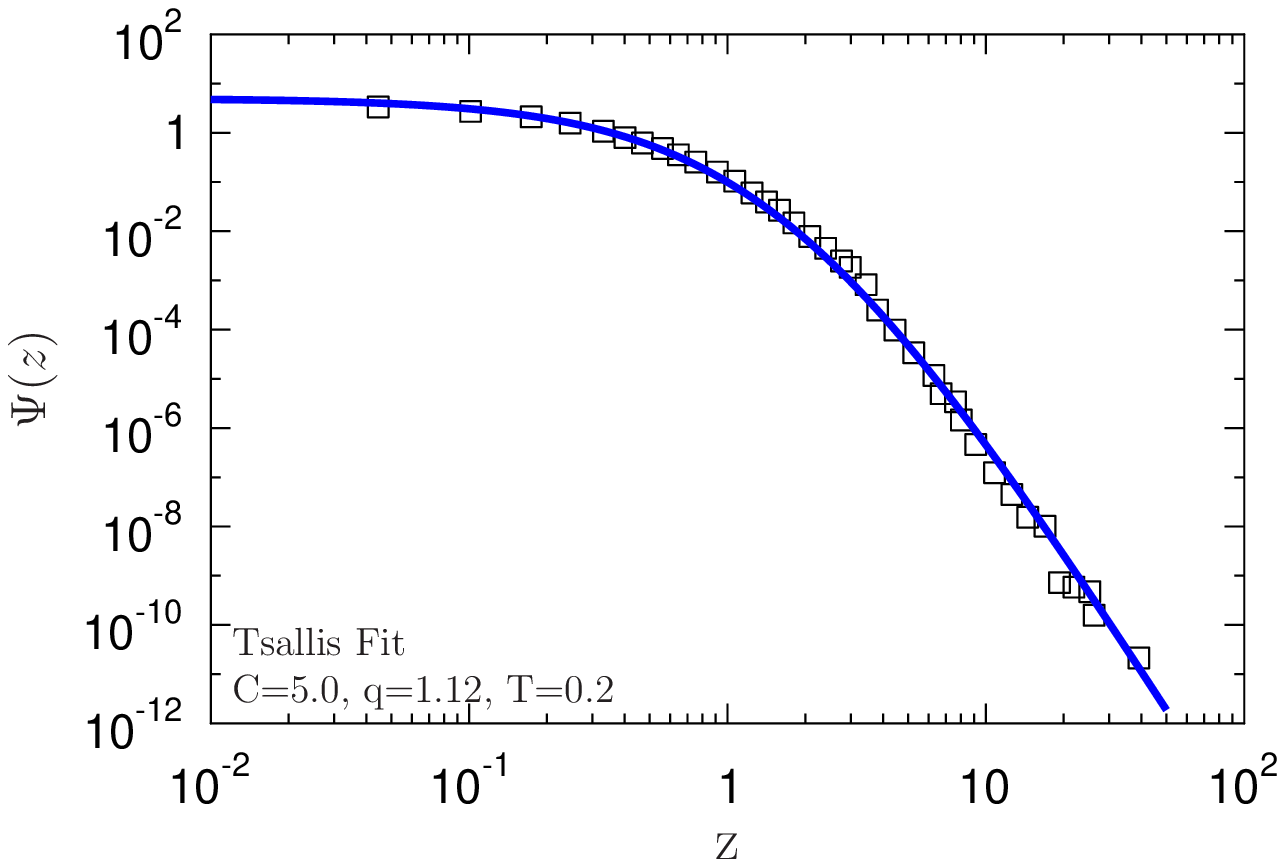}
\includegraphics[width=6.5cm]{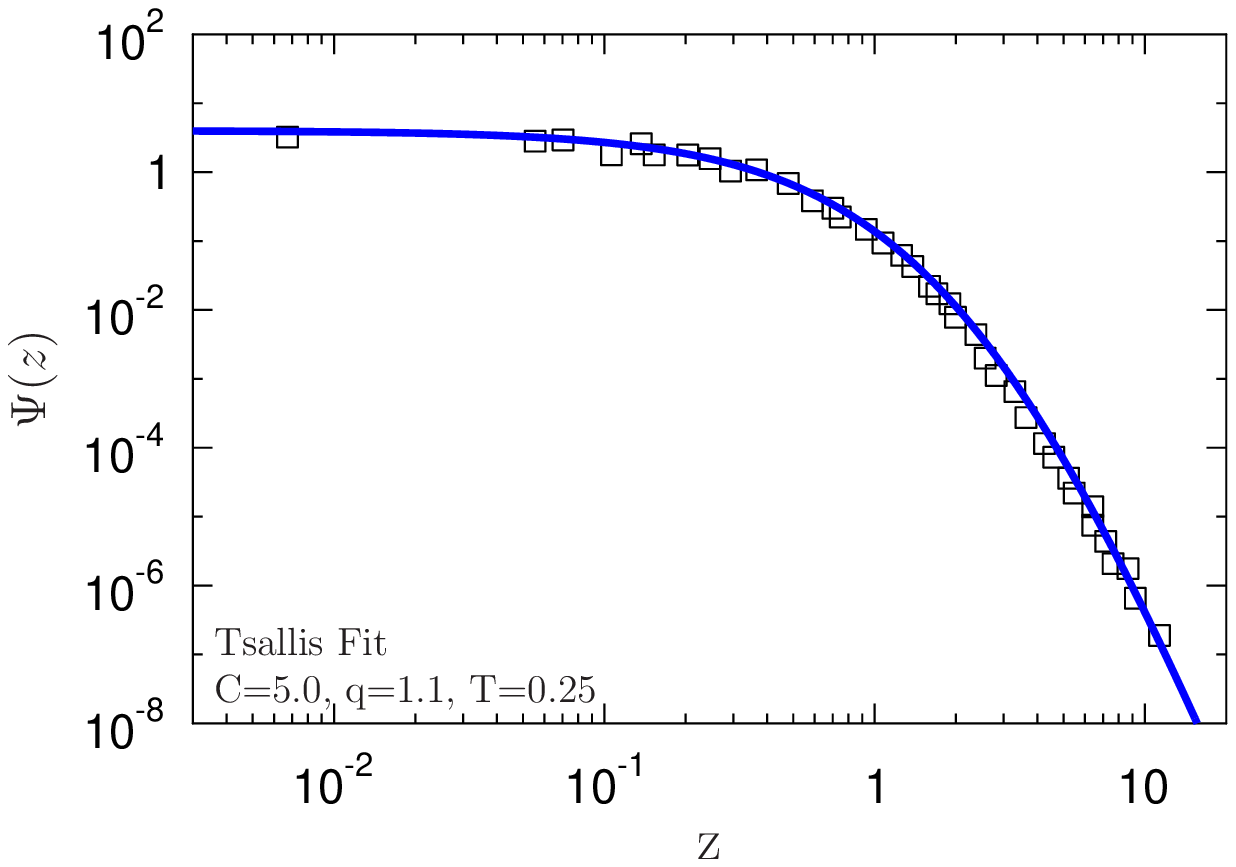}
\caption{(Color online) Tsallis fits to $z$-scaling plots
presented by Tokarev. Left panel: spectra of charged hadrons
produced in $pp$ collisions at energies $\sqrt{s} = 19-2360$ GeV
\cite{T_PTPS}; right panel: spectra of different particles
produced at $\sqrt{s} = 200$ GeV (from talk by Tokarev - these
proceedings).} \label{Fig2}
\end{figure}

It must be stressed at this point that essentially {\it all data}
observed in high energy production processes can be represented by
a universal scaling distribution, $\psi(z)$,
\cite{TZ,T_PTPS}\footnote{Where, in short: $\psi(z) = - \pi s
/\left[ (dN/d \eta) \sigma_{in} \right] J^{-1} E d^3\sigma/dp^3$,
with  $J$ being the corresponding Jacobian of transformations from
variables $\left\{ p_z, p_T \right\}$ to $\left\{z, \eta
\right\}$; $z = z_0\Omega^{-1}$ where $z_0 =
\sqrt{s_{T}}/\left[m\cdot\left(dN_{ch}/d\eta|_0\right)^c\right]$
and $\Omega^{-1}$ is the minimal resolution at which a constituent
subprocess can be singled out of the inclusive reaction,
$\sqrt{s_{T}}$ is the transverse kinetic energy of the subprocess
consumed on production of $m_1$ and $m_2$, $dN_{ch}/d\eta|_0$ is
the multiplicity density of charged particles at $\eta = 0$, $c$
is a parameter interpreted as a "specific heat" of created medium
and $m$ is is an arbitrary constant (usually fixed at the value of
nucleon mass).}. In Fig. \ref{Fig2} we demonstrate that these
curves can also be nicely reproduced by two-parameter Tsallis fits
$h_q(z)$. \vspace{-0.5cm}

\begin{figure}[htb]
\includegraphics[width=6.5cm]{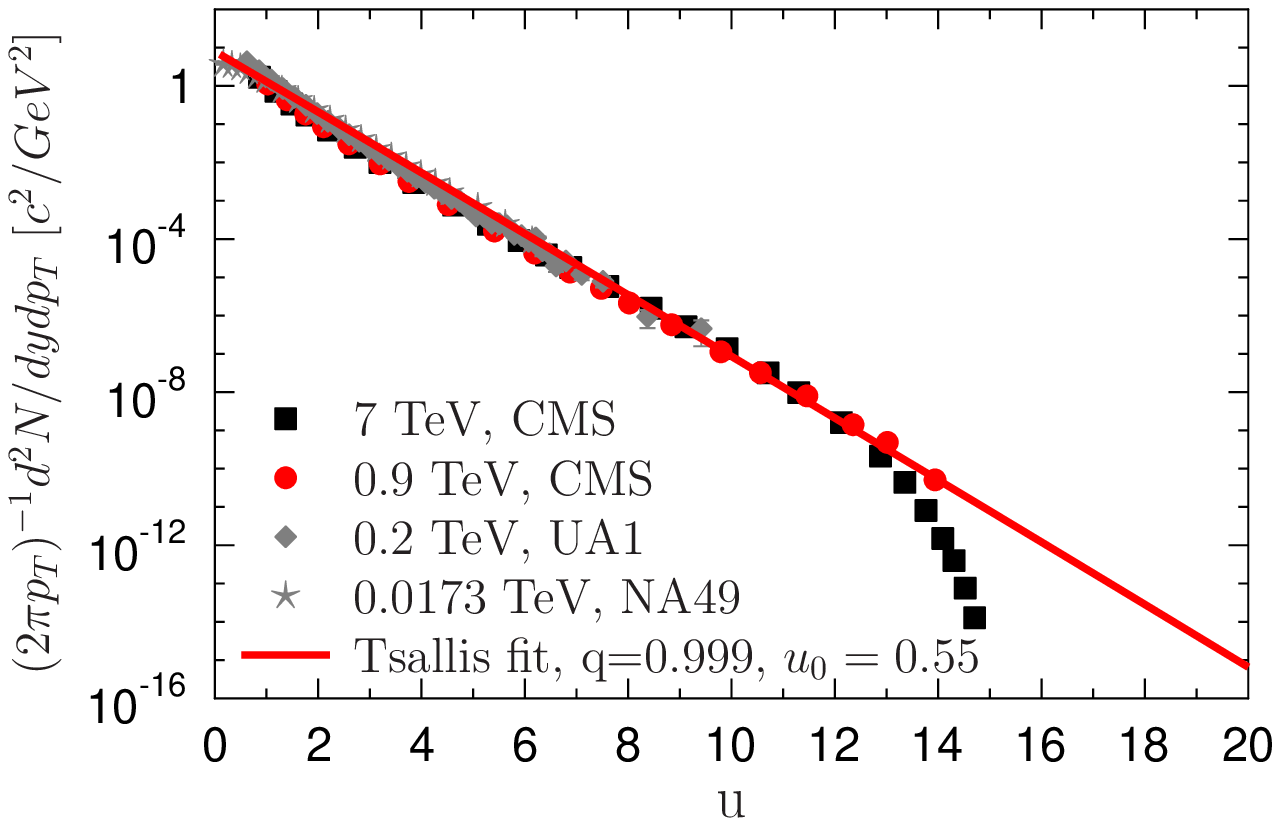}
\includegraphics[width=6.5cm]{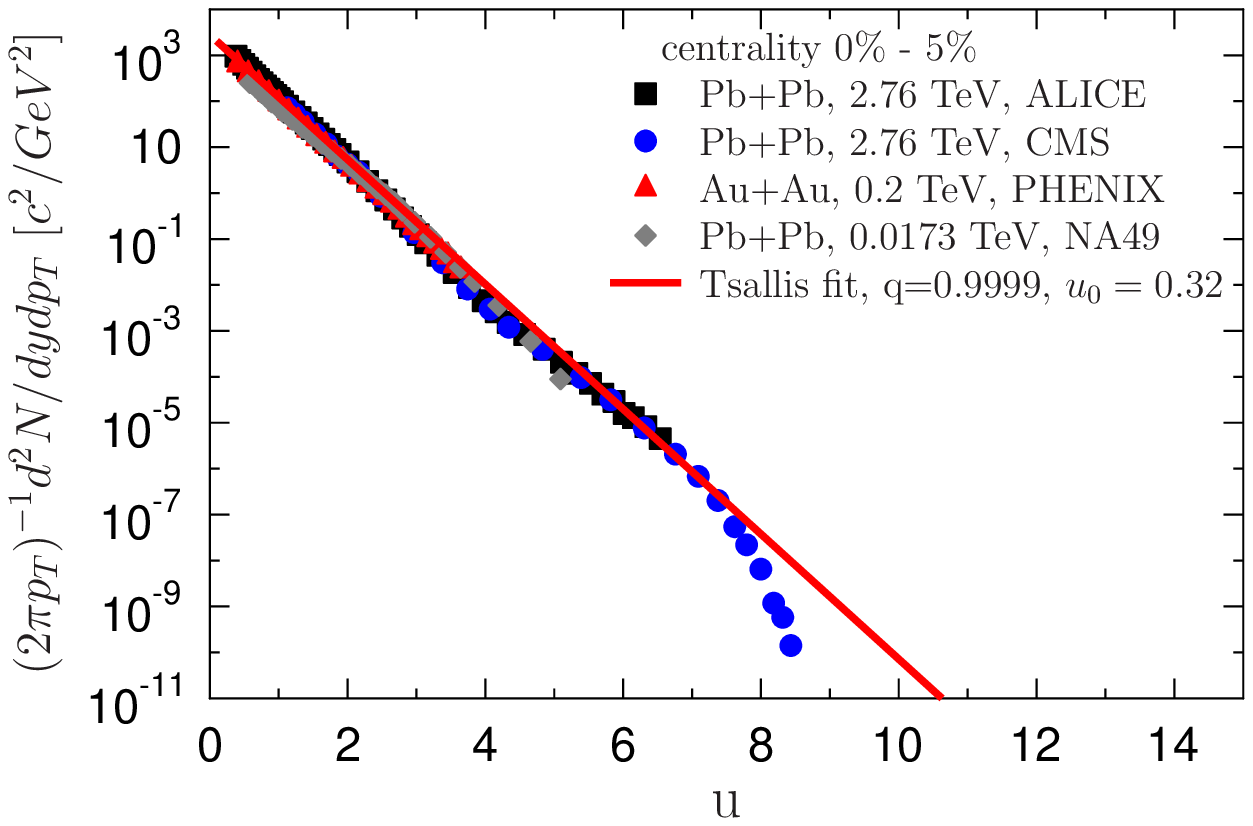}
\caption{(Color online) $p+p$, ($p+\bar{p}$) (left panel) and
central $A+A$ (right panel) data (from, respectively,
\cite{CMS,UA1,NA49} and \cite{ALICE,CMS_A,PHENIX_A,NA49_A}) for
transverse momentum distributions for different energies plotted
by using the scaling variable $u$.} \label{Fig3}
\end{figure}

\begin{figure}[htb]
\includegraphics[width=6.5cm]{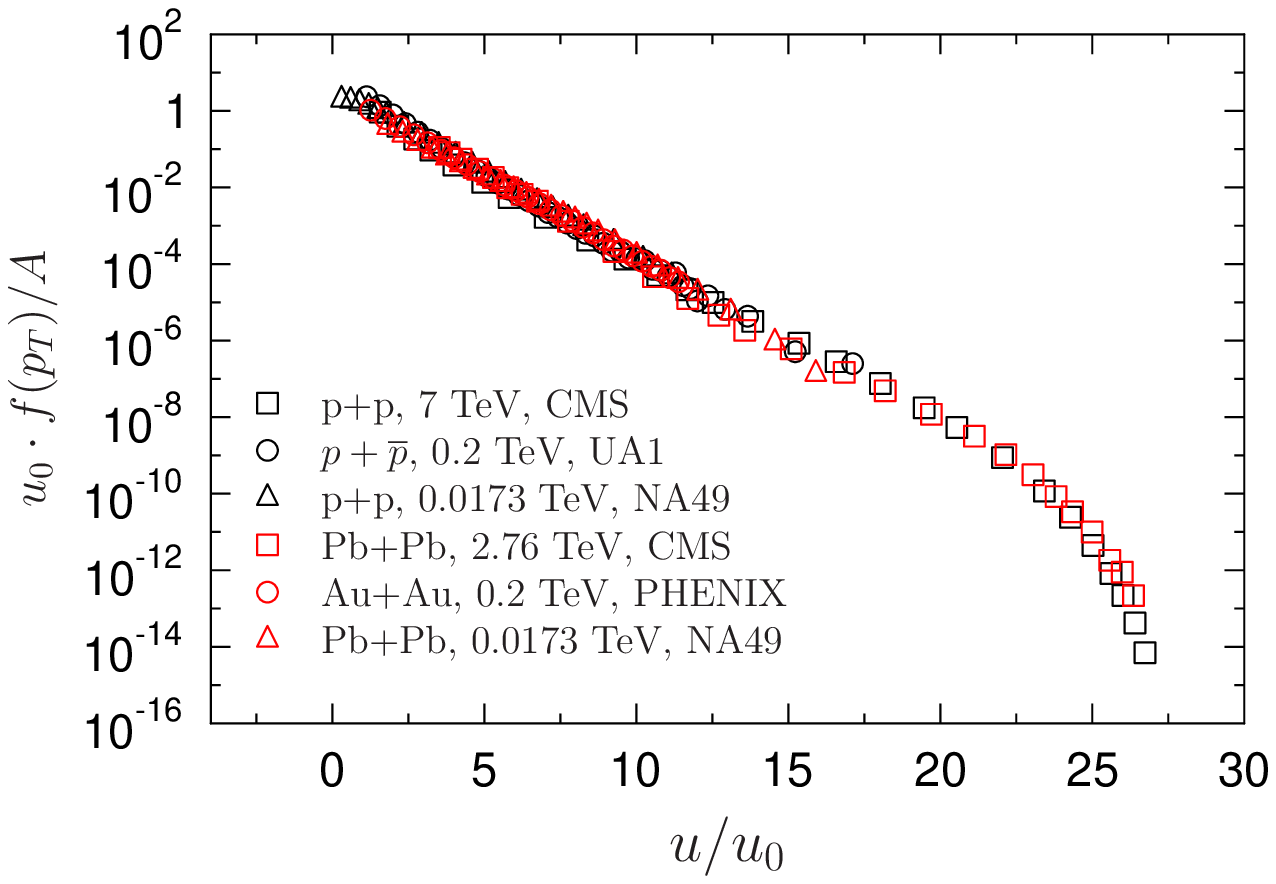}
\includegraphics[width=6.5cm]{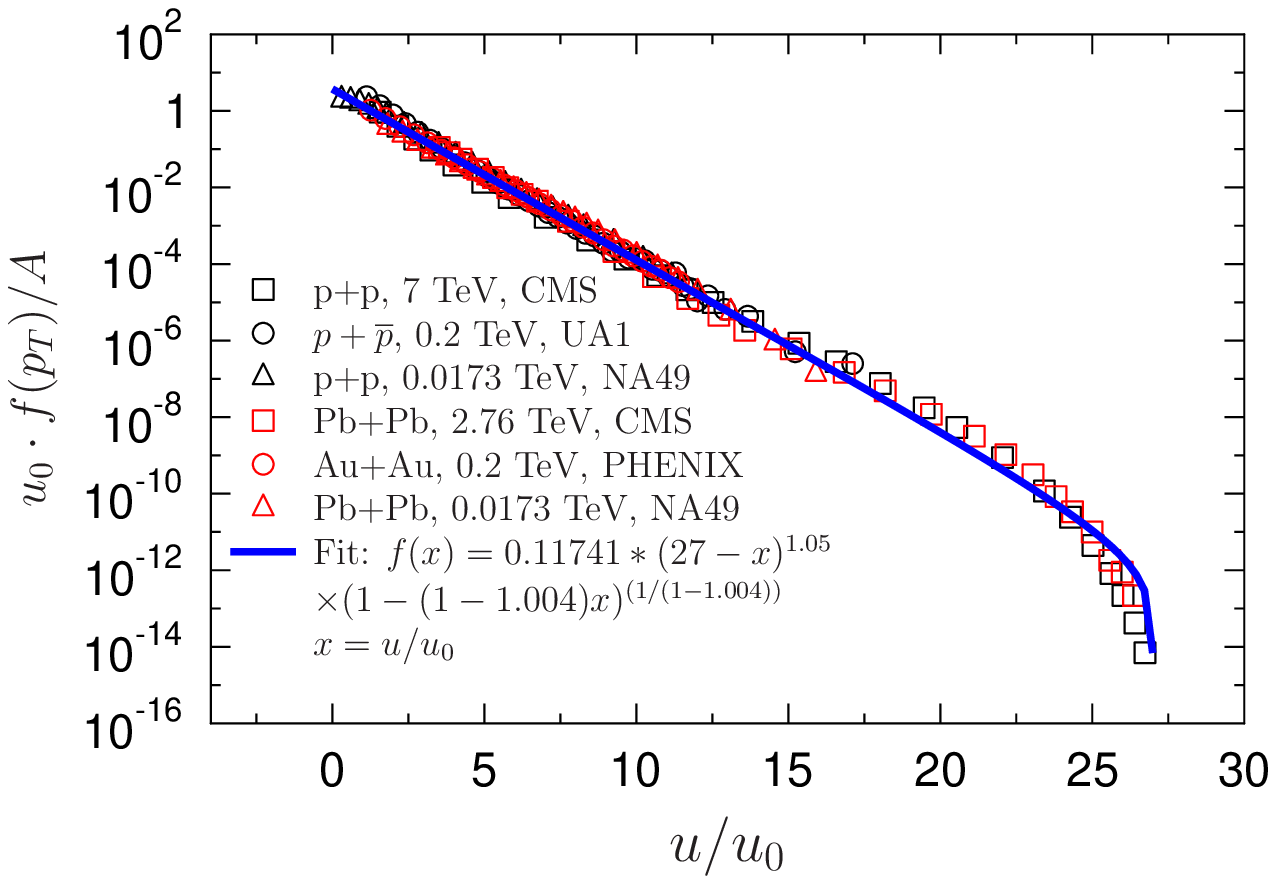}
\caption{(Color online) The same as in Fig. \ref{Fig3} but
multiplied by $u_0/A$ and plotted for $u/u_0$. Notice that curves
for $pp$ and $AA$ coincide (left panel) and can be fitted by the
same formula (right panel). The observed departure from an
exponent for large $u$ is connected with the effect of limitation
of phase space and "conditional probability" discussed in
\cite{cond}. } \label{Fig4}
\end{figure}

In this way we have reached our main point. Both
$h_q\left(p_T'\right)$ and $h_q(u)$ (not to mention $h_q(z)$)
depart from the usual connection of the nonextensive Tsallis
distribution with a thermodynamical approach (justified and
advocated in \cite{View}). Instead, as was already mentioned in
\cite{RWW}, we opt for some more dynamical source of $h_q$. As
shown there, our scaling seems to indicate close connections with
the so called {\it preferential attachment} (where
$T=T\left(p_T\right)$), seen in scale-free networks and dynamics
leading to it \cite{RWW}. In the case of variable $u$, this means
that the parameter $b$ can be positive or negative, depending on
circumstances. This in turn can result in growth or decrease of
$q$. In our case $q=1.172$, higher than obtained so far by using
the usual $h_q\left(p_T\right)$. The question then arises of
whether it is possible to get $h_{q\rightarrow 1}(u) ~\sim \exp
\left(-u/u_0\right)$ (by changing the sign of the parameter $b$ in
the definition of the variable $u$). In Fig. \ref{Fig3} we show a
comparison of data for $p+p$ and $A+A$ for such scaling. They are
different because of different $u_0$ and different normalizations
used. However, both curves coincide when multiplied by $u_0/A$ and
when plotted for $u/u_0$ (they become independent of $A$), cf.
Fig. \ref{Fig4}. They correspond to, respectively, $b(\sqrt{s}) =
-0.085+0.115(\sqrt{s})^{-0.2}$ for $p+p$ collisions and
$b(\sqrt{s}) = -0.052-0.0002 (\sqrt{s})^{0.7}$ for $A+A$
collisions. \vspace{-0.5cm}
\vspace{-0.5cm}

We close with some remarks concerning the variable $u$. As already
said, by switching to this variable we depart from the usual
thermal-like description of such processes because $u$ contains
some dynamical input, for example:
\begin{itemize}
\item One can write $u/u_0 = p_T/T_{eff}$ where $T_{eff}$ is an
effective temperature: $T_{eff} = T_0 + T_v\left( p_T\right)$ with
$T_0 = u_0\cdot \langle p_T\rangle$ and $T_v = - b\cdot u_0\cdot
p_T$. Such $T_{eff}$ could be related to the possible $p_T$
transfer, additional to that resulting from a hard collision,
perhaps proceeding by a kind of multiple scattering
process\footnote{Similar, in a sense, to that proposed on a
different occasion in \cite{multiple}.}.

\item  $T_{eff}$ also occurs in a description of the growth of the
so called complex free networks which can then be applied to
hadronic production (\cite{RWW}, cf. also \cite{nets} and
references therein).
\end{itemize}

Lets elaborate some more on the second proposition. Here one
associates $p_T$ with the number of links in the quark-gluonic
network assumed to be formed in the hadronization process. In this
case their actual original energy-momentum distributions would be
of secondary importance, since, because of their mutual
interactions, they connect to each other and this process of
connection has its distinctive dynamical consequences. One can
think of such a process in the following way: We start with some
initial state consisting of a number $n_0$ of already existing
($q\bar{q}$) pairs (identified with vertices in the network) and
we add to them, in each consecutive time step, another vertex (a
new ($q\bar{q}$) pair), which can have $k_0$ possible connections
(links in network language) to the old state. Quarks are dressed
by interaction with surrounding gluons and therefore "excited" and
each quark interacts with $k$ other quarks (has $k$ links).
Assuming that the "excitation" of a quark is proportional to the
number of links $k$ (which is proportional to the number of gluons
participating in "excitation", i.e., existing in the vicinity of a
given quark), the chances to interact with a given quark grow with
the number of links attached to it, $k$. The new links will be
preferentially attached to quarks with $k$. This corresponds to
building up a so called preferential network, which evolves due to
the occurrence of new ($q\bar{q}$) pairs from decaying gluons.
Such networks always result in a power-like behavior of suitable
variables, in our case in $p_T$.

\vspace*{0.3cm} \centerline{\bf Acknowledgment}

This research  was supported in part by the National Science
Center (NCN) under contract DEC- 2011/03/B/ST2/02617 (MR and ZW)
and by the Ministry of Science and Higher Education under contract
DPN/N97/CERN/2009 (GW). We would like to warmly thank Dr Eryk
Infeld for reading this manuscript.

\end{document}